\documentclass[twoside,fleqn,pdftex]{article}
\usepackage{espcrc2}
\usepackage{epsfig,macros}
\usepackage{amssymb,amsmath}
\title{%
Towards precision heavy flavour physics from lattice QCD%
}

\author{%
Jochen Heitger\address{%
Westf\"alische Wilhelms-Universit\"at M\"unster, 
Institut f\"ur Theoretische Physik,\\
Wilhelm-Klemm-Stra{\ss}e~9, D-48149 M\"unster, Germany}
(ALPHA Collaboration)
}

\begin{document}


\thispagestyle{empty}

\begin{flushright}
MS-TP-11-01
\end{flushright}

\vspace{2.0cm}
\begin{center}
\boldmath
{\large\bf%
Towards precision heavy flavour physics from lattice QCD
}
\unboldmath
\end{center}

\vspace{1.0cm}
\begin{center}
Jochen Heitger\\[0.2cm]
{\sl%
Westf\"alische Wilhelms-Universit\"at M\"unster,
Institut f\"ur Theoretische Physik\\
Wilhelm-Klemm-Stra{\ss}e~9, D-48149 M\"unster, Germany
}\\[-7.0cm]
\vbox{%
\includegraphics[width=0.375\textwidth]{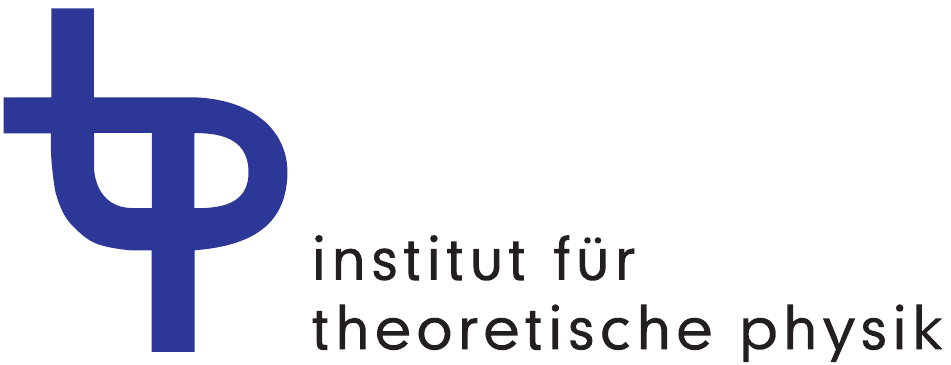}
\hspace{1.0cm}
\includegraphics[width=0.125\textwidth]{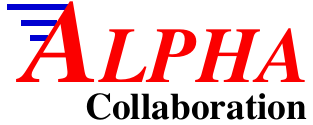}
}
\end{center}

\vspace{1.0cm}
\begin{center}
{\bf Abstract}
\end{center}

{\small%
\noindent%
I convey an idea of the significant recent progress, which opens up good 
perspectives for high-precision ab-initio computations in heavy flavour 
physics based on lattice QCD.
This report focuses on the strategy and the challenges of fully 
non-perturbative investigations in the B-meson sector, where the b-quark is 
treated within an effective theory, as followed by the ALPHA Collaboration.
As an application, I outline its use to determine the b-quark mass and 
summarize the status of our ongoing project in the two dynamical flavour 
theory.
}

\vspace{1.0cm}
\begin{center}
Invited talk at the\\[0.2cm]
{\sl%
Third Workshop on Theory, Phenomenology and Experiments in Heavy Flavour 
Physics\\ 
July 5 -- 7, 2010, in Anacapri, Capri, Italy
}\\[0.2cm]
To appear in the Proceedings (Nucl.~Phys.~B~Proc.~Suppl.)
\end{center}

\vspace{1.0cm}
\begin{center}
\noindent%
November 2010
\end{center}
\vspace{2.0cm}
\vfill

\newpage
\thispagestyle{empty}
\vbox{}
\newpage
 
\setcounter{page}{1}


\begin{abstract}
I convey an idea of the significant recent progress, which opens up good 
perspectives for high-precision ab-initio computations in heavy flavour 
physics based on lattice QCD.
This report focuses on the strategy and the challenges of fully 
non-perturbative investigations in the B-meson sector, where the b-quark is 
treated within an effective theory, as followed by the ALPHA Collaboration.
As an application, I outline its use to determine the b-quark mass and 
summarize the status of our ongoing project in the two dynamical flavour 
theory.
\end{abstract}

\maketitle
%
\section{B-physics and lattice QCD}
\label{Sec_Bphys}
For the plenty of beautiful results from recent and current B-physics 
experiments~\cite{HFAG,Altarelli:2009ec} --- as well as from what is to be
expected from LHC ---, to lead to feasible precision tests of the Standard 
Model and trials of several New Physics scenarios, requires the knowledge of 
QCD matrix elements for their interpretation in terms of parameters of the 
Standard Model and its possible extensions.  
Unfortunately, the uncertainty on the theoretical side in this interplay of 
experiment and theory in flavour physics predominantly originates from 
hardly computable long-distance effects of the strong interaction that 
confines quarks and gluons within hadrons.
This potentially limits the impact of future experimental measurements on 
New Physics models and motivates calculations in lattice QCD, which is a 
powerful approach to reach a few-\% theoretical error on those 
non-perturbative hadronic contributions.

Still, some care is needed to obtain reliable results for b-quark physics
from a Monte Carlo evaluation of the discretized Euclidean path integral.
One has to keep under control simultaneously the finite-size effects and, 
particularly, the discretization effects, since the lattice spacing should 
not be larger than the Compton length of the b-quark. 
In practice, it is not possible to control both effects by brute force 
numerical simulations such that dedicated methods have to be devised.

While the numerical computations in lattice QCD necessarily involve 
approximations, one of the key features of the lattice approach is that all 
approximations can be systematically improved. 
For an overview of the different formulations of heavy quarks on the lattice
that have been proposed in the literature and are being used today, and of 
results from the field of heavy flavour physics, which reflect some of these 
improvements by the small error bars quoted for many quantities, I refer to 
the reviews of past Lattice 
Conferences~\cite{lat07:michele,lat08:bphys,lat09:bphys,lat09:ckmphys,lat10:HQPjochen}
and references therein.
\subsection{Challenges}
\label{Sec_Bphys_chall}
\begin{figure}[htb]
\begin{center}
\vspace{-0.75cm}
\includegraphics[width=0.4625\textwidth]{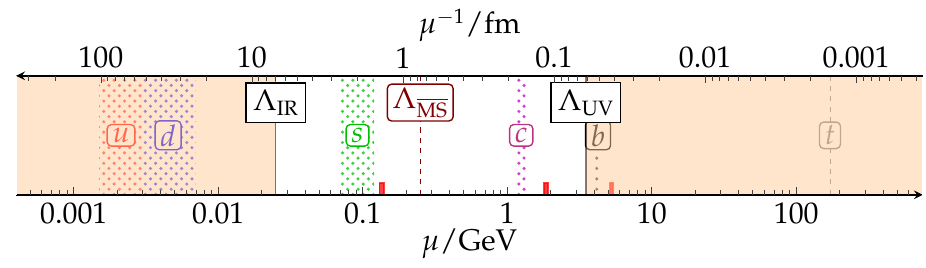}
\vspace{-1.5cm}
\caption{\sl%
Large range of energy ($\mu$) scales in lattice QCD, where shaded areas 
refer to quark mass values (in the $\MSbar$ scheme) quoted by the Particle 
Data Group \cite{PDBook}.
Red marks indicate the pion, the D- and the B-meson mass. 
}\label{fig:massscales}
\end{center}
\vspace{-0.75cm}
\end{figure}

Among the various considerable challenges one faces in an actual lattice
QCD calculation on the theoretical and technical levels, let us only 
highlight the multi-scale problem, which is also particularly relevant in 
view of B-physics applications.
This is illustrated in \Fig{fig:massscales}.
%
%
There are many disparate physical scales to be covered simultaneously, 
ranging from the lightest hadron mass of $\mpi\approx 140\,\MeV$ over
$\mD\approx 2\,\GeV$ to $\mB\approx5\,\GeV$, plus the ultraviolet cutoff of 
$\Lambda_{\rm UV}=a^{-1}$ of the lattice discretization that has to be large 
compared to all physical energy scales for the discretized theory to be an 
approximation to the continuum one.
Moreover, the finiteness of the linear extent of space-time, $L$, in a 
numerical treatment entails an infrared cutoff $\Lambda_{\rm IR}=L^{-1}$ so 
that the following scale hierarchy is met:
\[
\Lambda_{\rm IR}=L^{-1}
\,\ll\, 
\mpi,\ldots,\mD,\mB
\,\ll\,
a^{-1}=\Lambda_{\rm UV}\,. 
\]
This implies $L\gtrsim 4/\mpi\approx 6\,\Fm$ to suppress finite-size effects 
in the light quark sector and $a\lesssim 1/(2\mD)\approx 0.05\,\Fm$ to still
properly resolve the propagation of a c-quark in the heavy sector.
Lattices with $L/a \gtrsim 120$ sites in each direction would thus be 
needed to satisfy these constraints, and since the scale of hadrons with
b-quarks was not even included to arrive at this figure, it is obvious that 
the b-quark mass scale has to be separated from the others in a 
theoretically sound way before simulating the theory.
In \Sect{Sec_npHQET} I briefly describe, how this is achieved by recoursing 
to an effective theory for the b-quark.

Another non-trivial task is the renormalization of QCD operators composed 
of quark and gluon fields, which appear in the effective weak Hamiltonian, 
valid at energies far below the electroweak scale.
Besides perturbation theory (see, e.g., \cite{LPT:stefano}), powerful 
non-perturbative approaches have been developed 
(and reviewed, e.g., in \cite{reviews:NPRrainer_nara}), and I will come 
back to the non-perturbative subtraction of power-law divergences in the 
context of the effective theory for the b-quark later.
\subsection{Perspectives}
\label{Sec_Bphys_persp}
As for the challenges with light quarks, it should only be noted that the 
condition $L\gtrsim 6\,\Fm$ may be relaxed by simulating at unphysically 
large pion masses, combined with a subsequent extrapolation guided by chiral 
perturbation theory~\cite{chPT:GaLe1} and its lattice-specific refinements.

Regarding the algorithmic side of a lattice QCD simulation, the Hybrid Monte 
Carlo \cite{hmc:orig} (HMC) as the first exact and still state-of-the-art 
algorithm has received considerable improvements by multiple time-scale 
integration schemes \cite{hmc:mtsi1,hmc:mtsi2}, the Hasenbusch trick of 
mass-preconditioning \cite{hmc:hasenb1,hmc:hasenb2}, supplemented by a
sensible tuning of the algorithm's parameters \cite{Nf2SF:algo}, and domain 
decomposition (DD) applied to 
QCD \cite{ddhmc:luescher1,ddhmc:luescher2,ddhmc:luescher3}, just to name
a few.
In addition, low-mode deflation \cite{deflat:luescher2} (together with
chronological inverters \cite{Brower:1995vx}) has led to a substantial 
reduction of the critical slowing down with the quark mass in the DD-HMC.

Finally, in parallel to the continuous increase of computer speed
(at an exponential rate) over the last 25 years and the recent investments 
into high performance computing at many places of the world, the
Coordinated Lattice Simulations \cite{community:CLS} (CLS) initiative is a 
community effort to bring together the human and computer resources of 
several teams in Europe interested in lattice QCD.
The present goal are large-volume simulations with $\nf=2$ dynamical quarks,
using the rather simple $\Or(a)$ improved Wilson action to profit from the 
above algorithmic developments such as DD-HMC, and lattice spacings 
$a=(0.08-0.05)\,\Fm$, sizes $L=(2-4)\,\Fm$ and pion masses down to 
$\mpi=250\,\MeV$, which altogether help to diminish systematic and 
statistical errors.
Amongst others, the B-physics programme outlined here is investigated on
these lattices. 
\section{Non-perturbative HQET}
\label{Sec_npHQET}
Heavy Quark Effective Theory (HQET) at zero velocity on the 
lattice~\cite{stat:eichhill1} offers a reliable solution to the problem of 
dealing with the two disparate intrinsic scales encountered in heavy-light 
systems involving the b-quark, i.e., the lattice spacing $a$, which has to 
be much smaller than $1/m_{\rm b}$ to allow for a fine enough resolution of 
the states in question, and the linear extent $L$ of the lattice volume, 
which has to be large enough for finite-size effects to be under control
(\Fig{fig:massscales}).

Since the heavy quark mass ($\mb$) is much larger than the other scales 
such as its 3--momentum or $\lQCD\sim 500\,\MeV$, HQET relies upon a 
systematic expansion of the QCD action and correlation functions in inverse 
powers of the heavy quark mass around the static limit ($\mb\to\infty$).
The lattice HQET action $S_{\rm HQET}$ at $\Or(1/\mb)$ reads:
\[
a^4{\T \sum_x}\heavyb\left\{
D_{0}+\dmstat-\omkin\vecD^2-\omspin\vecsigma\vecB
\right\}\heavy\,,
\label{shqet}
\]
with $\heavy$ satisfying $P_+\heavy=\heavy$, $P_+={{1+\gamma_0}\over{2}}$,
and the parameters $\omega_{\rm kin}$ and $\omega_{\rm spin}$ being formally 
$\Or(1/\mb)$.
At leading order (static limit), where the heavy quark acts only as a 
static colour source and the light quarks are independent of the heavy 
quark's flavour and spin, the theory is expected to have $\sim 10\%$ 
precision, while this reduces to $\sim 1\%$ at $\Or(1/\mb)$ representing the 
interactions due to the motion and the spin of the heavy quark. 
As crucial advantage (e.g., over NRQCD), HQET treats the 
$1/\mb$--corrections to the static theory as space-time insertions in 
correlations functions.
For correlation functions of some multi-local fields $\op{}$ and up to 
$1/m_{\rm b}$--corrections to the operator itself (irrelevant when spectral 
quantities are considered), this means
\bean
\langle\op{}\rangle=
\langle\op{}\rangle_{\mrm{stat}}+a^4\sum_x\Big\{
& &
\hspace{-0.75cm}
\omkin\langle\op{}\Okin(x)\rangle_{\mrm{stat}}\\
& &
\hspace{-0.75cm}
+\,\omspin\langle\op{}\Ospin(x)\rangle_{\mrm{stat}}
\Big\}\,,
\eean
where $\langle\op{}\rangle_{\rm stat}$ denotes the expectation value in the 
static approximation and $\Okin$ and $\Ospin$ are given by 
$\heavyb\vecD^2\heavy$ and $\heavyb\vecsigma\vecB\heavy$.
In this way, HQET at a given order is (power-counting) renormalizable and
its continuum limit well defined, once the mass counterterm $\dmstat$ and
the coefficients $\omkin$ and $\omspin$ are fixed non-perturbatively by a 
matching to QCD.

Still, for lattice HQET and its numerical applications to lead to precise 
results with controlled systematic errors in practice, two shortcomings had 
to be left behind first.

1.) The exponential growth of the noise-to-signal ratio in static-light 
correlators, which is overcome by a clever modification of the Eichten-Hill 
discretization of the static action~\cite{HQET:statprec}.

2.) As in HQET mixings among operators~of different dimensions occur, the
power-divergent additive mass renormalization $\dmstat\sim g_0^2/a$
already affects its leading order.
Unless HQET is renormalized non-perturbatively~\cite{Maiani:1992az}, this 
divergence --- and those $\sim g_0^2/a^{2}$ arising at $\Or(1/\mb)$ --- 
imply that the continuum limit does not exist owing to a 
remainder, which, at any finite perturbative 
order~\cite{mbstat:dm_MaSa,mbstat:dm_DirScor}, diverges as $a\to 0$.
A general solution to this theoretically serious problem was worked out and 
implemented for a determination of the b-quark's mass in the static and 
quenched approximations as a test case~\cite{HQET:pap1}.
It is based on a \emph{non-perturbative matching of HQET and QCD in 
finite volume}.
Applications of this strategy to the determination of the b-quark mass and
(a subset of all) HQET parameters at 
$\Or(1/\mb)$ \cite{HQET:mb1m,HQET:param1m}, to a study of the
${\rm B}_{\rm s}$-meson spectrum \cite{HQET:msplit} and to a computation of 
the ${\rm B}_{\rm s}$-meson decay constant \cite{HQET:fb1m} were realized 
in the quenched approximation by our collaboration and have been extended 
to the more realistic $\nf=2$ 
situation~\cite{lat07:hqetNf2,lat08:hqettests,impr:babp_nf2,lat10:hqetNf2}.
\section{The b-quark mass via HQET at $\Or(1/\mb)$}
\label{Sec_appl}
We first note~\cite{reviews:NPHQETrainer} that in order not to spoil the 
asymptotic convergence of the series, the matching must be done 
non-perturbatively --- at least for the leading, static piece --- as soon as 
the $1/\mb$--corrections are included, since
as $\mb\to\infty$ the \emph{perturbative} truncation error from the 
matching coefficient of the static term becomes much larger than the power 
corrections $\sim\lQCD/\mb$ of the HQET expansion.

%
\begin{figure}[htb]
\begin{center}
\vspace{-0.75cm}
\includegraphics[width=0.4625\textwidth]{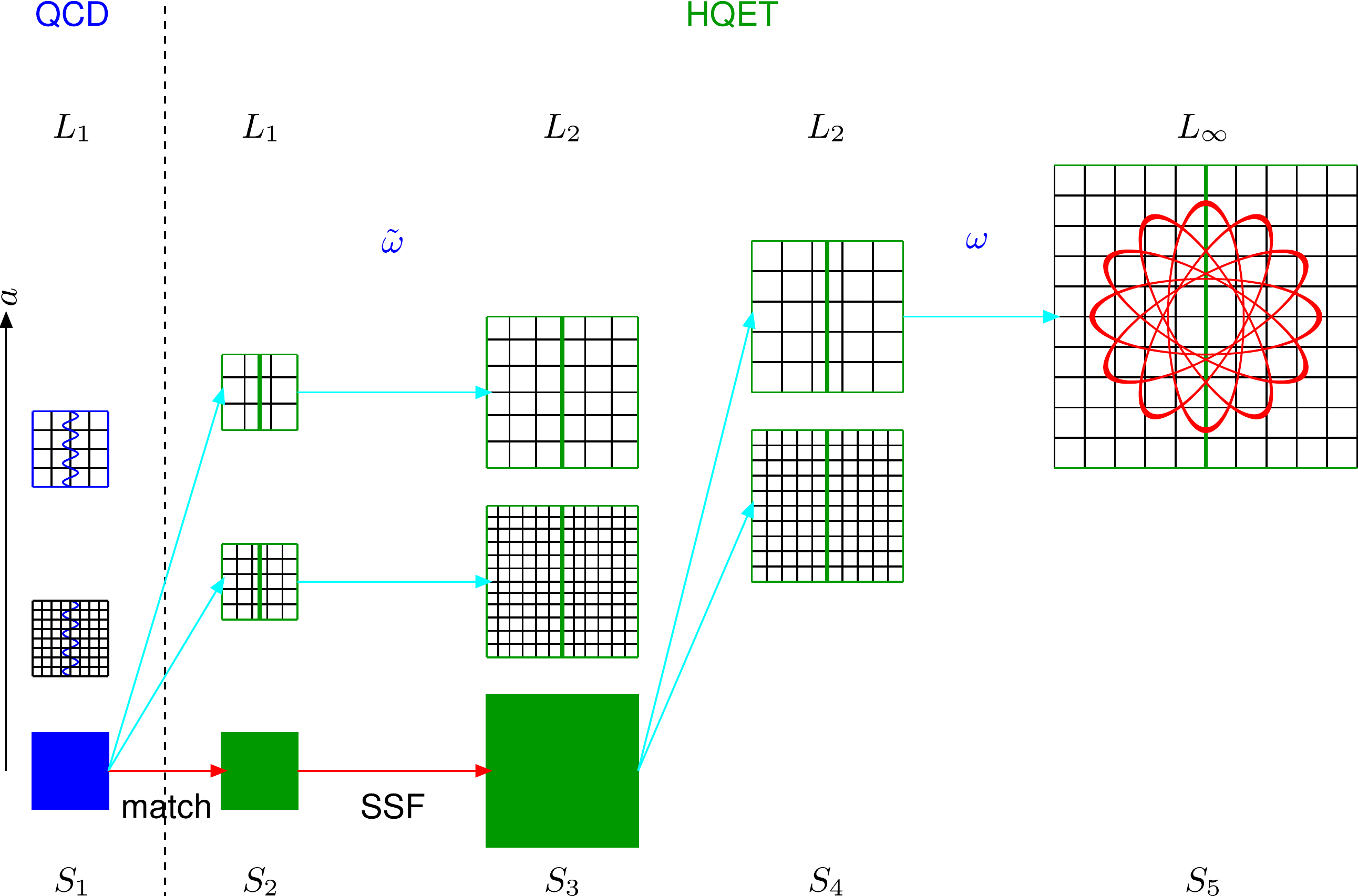}
\vspace{-1.25cm}
\caption{\sl%
Idea of lattice HQET computations via a non-perturbative determination of 
HQET parameters from small-volume QCD simulations.
For each fixed $L_i$, the steps are repeated at smaller $a$ to reach the 
continuum limit.
}\label{fig:strat}
\end{center}
\vspace{-0.75cm}
\end{figure}
%
In the framework introduced in~\cite{HQET:pap1}, matching and 
renormalization are performed simultaneously \emph{and} non-perturbatively.
Let us here explain the general strategy, illustrated in~\Fig{fig:strat}, 
for the sample application of calculating the b-quark mass.
$S_1$:
Starting from a finite volume with $L_1\approx 0.5\,\Fm$, one chooses 
lattice spacings $a$ sufficiently smaller than $1/\mb$ such that the 
b-quark propagates correctly up to controllable discretization errors of 
order $a^2$. 
The relation between the renormalization group invariant (RGI) and the bare 
mass in QCD being known, suitable finite-volume observables 
$\Phi_k(L_1,M_{\rm h})$ can be calculated as a function of the RGI heavy 
quark mass, $M_{\rm h}$, and extrapolated to the continuum limit. 
$S_2$:
Next, the power-divergent subtractions are performed non-perturbatively by 
a set of matching conditions, in which the results obtained for $\Phi_k$ 
are equated to their representation in HQET.
At the same physical value of $L_1$ but for resolutions $L_1/a=\rmO(10)$, 
the previously computed heavy-quark mass dependence of 
$\Phi_k(L_1,M_{\rm h}) $ in finite-volume QCD may be exploited to determine 
the bare parameters of HQET for $a\approx(0.025-0.05)\,\Fm$.
$S_3$:
To evolve the HQET observables to large volumes, where contact with some 
physical input from experiment can be made, one also computes them at these 
lattice spacings in a larger volume, $L_2=2L_1$.
The resulting relation between $\Phi_k(L_1)$ and $\Phi_k(L_2)$ is encoded in
associated step scaling functions denoted as $\sigma_k$.
$S_4,S_5$:
By using the knowledge of $\Phi_k(L_2,M_{\rm h})$ one fixes the bare 
parameters of the effective theory for $a\approx(0.05-0.1)\,\Fm$ so that a 
connection to lattice spacings is established, where large-volume 
observables, such as the B-meson mass or decay constant, can be calculated. 
This sequence of steps yields an expression of $\mB$, the physical input, as 
a function of $M_{\rm h}$ via the quark mass dependence of 
$\Phi_k(L_1,M_{\rm h})$, which eventually is inverted to arrive at the 
desired value of the RGI b-mass within HQET.
The whole construction is such that the continuum limit can be taken for 
all pieces. 
\subsection{Computation of HQET parameters}
\label{Sec_appl_param}
Following the strategy sketched above and applied to the quenched case 
in~\cite{HQET:param1m}, the determination of the parameters of the HQET 
Lagrangian and of the time component of the isovector axial current is
performed within the Schr\"odinger functional, i.e., QCD with Dirichlet 
boundary conditions in time and periodic ones in space, where suitable 
matching observables $\Phi_k$, such as finite-volume meson energies and 
matrix elements, can be readily defined.
Relativistic quarks are simulated as clover-improved Wilson fermions with 
$\nf=2$ dynamical quarks; for the static quark we use the so-called HYP1/2 
actions~\cite{HQET:statprec}.

\begin{figure}[htb]
\begin{center}
\vspace{-0.75cm}
\includegraphics[width=0.23\textwidth]{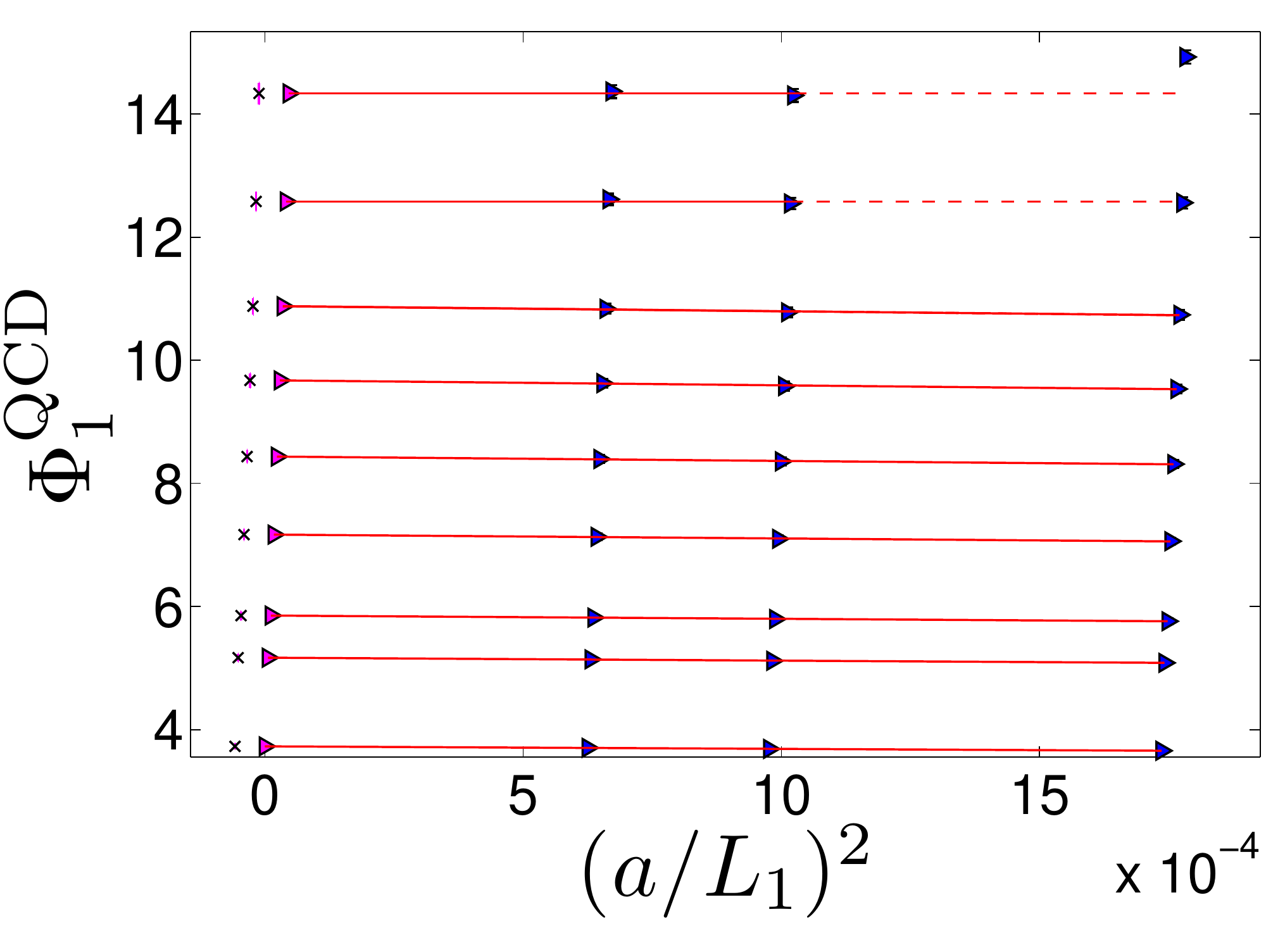}
\includegraphics[width=0.23\textwidth]{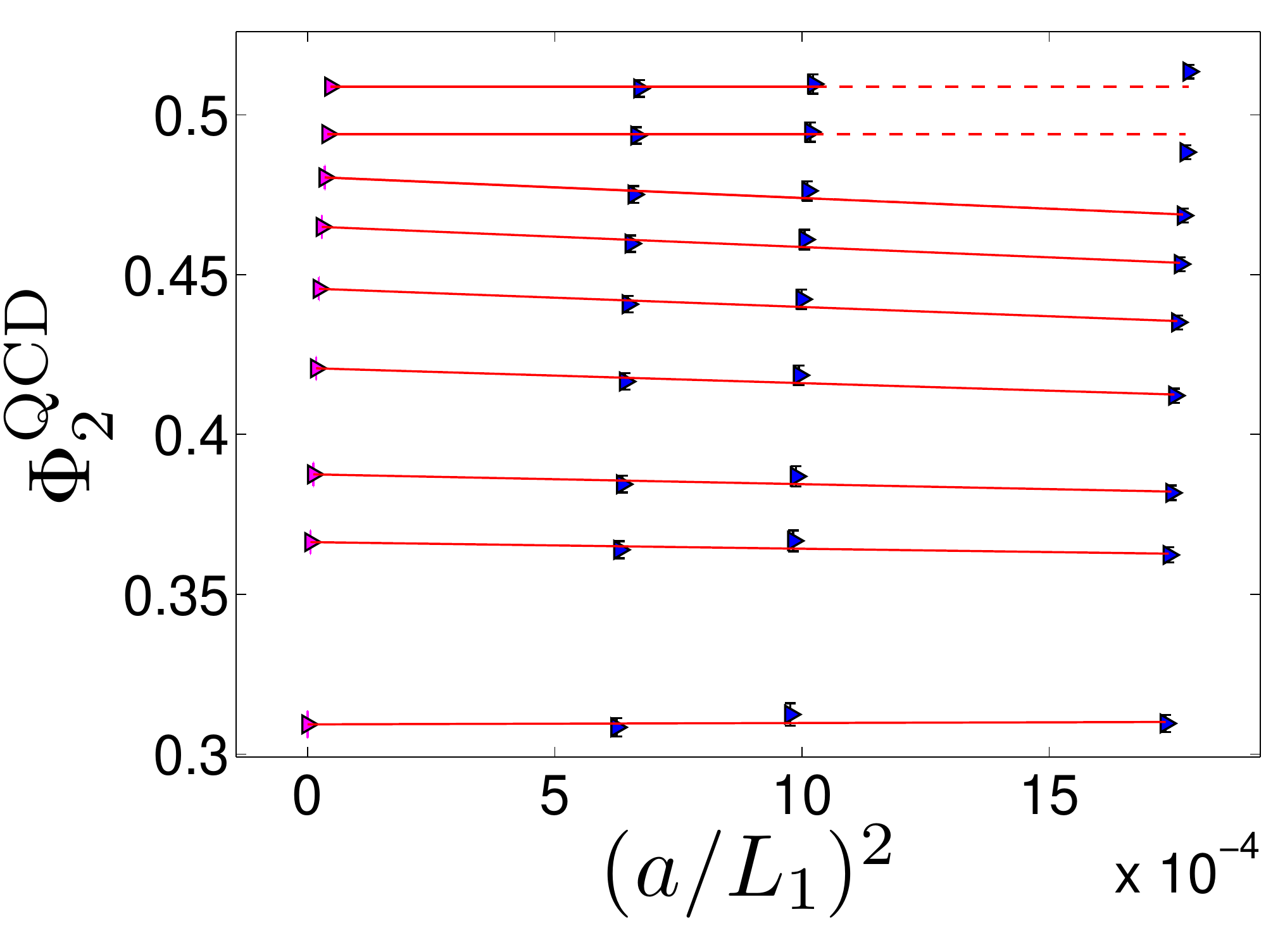}
\vspace{-1.25cm}
\caption{\sl%
Continuum extrapolation of the finite-volume observables $\Phi_1$ and 
$\Phi_2$, where for $\Phi_1$ we have included the error (cross on the left)
stemming from the renormalization of the quark mass.
}\label{fig:Phi12L1}
\vspace{-0.75cm}
\end{center}
\end{figure}

We introduce observables $\Phi_{k=1,\ldots,5}$ casted into a vector 
$\Phi\equiv\Phi^{\rm QCD}$, where in the continuum and large volume limits, 
the first two are proportional to the meson mass and to the logarithm of the 
decay constant, respectively, while $\Phi_3$ is used to fix the counterterm 
of the axial current and $\Phi_{4,5}$ for the determination of the kinetic 
and magnetic terms in $S_{\rm HQET}$.
The continuum extrapolations of $\Phi_{1,2}$ in the small QCD volume 
($L_1\approx 0.5\,\Fm$, $S_1$ in \Fig{fig:strat}), for nine values 
$\Mh\equiv M$ of the RGI heavy quark mass from the charm to beyond the 
bottom region~\cite{impr:babp_nf2}, are shown in \Fig{fig:Phi12L1}.
%
%

\begin{figure}[htb]
\begin{center}
\vspace{-0.75cm}
\includegraphics[width=0.23\textwidth]{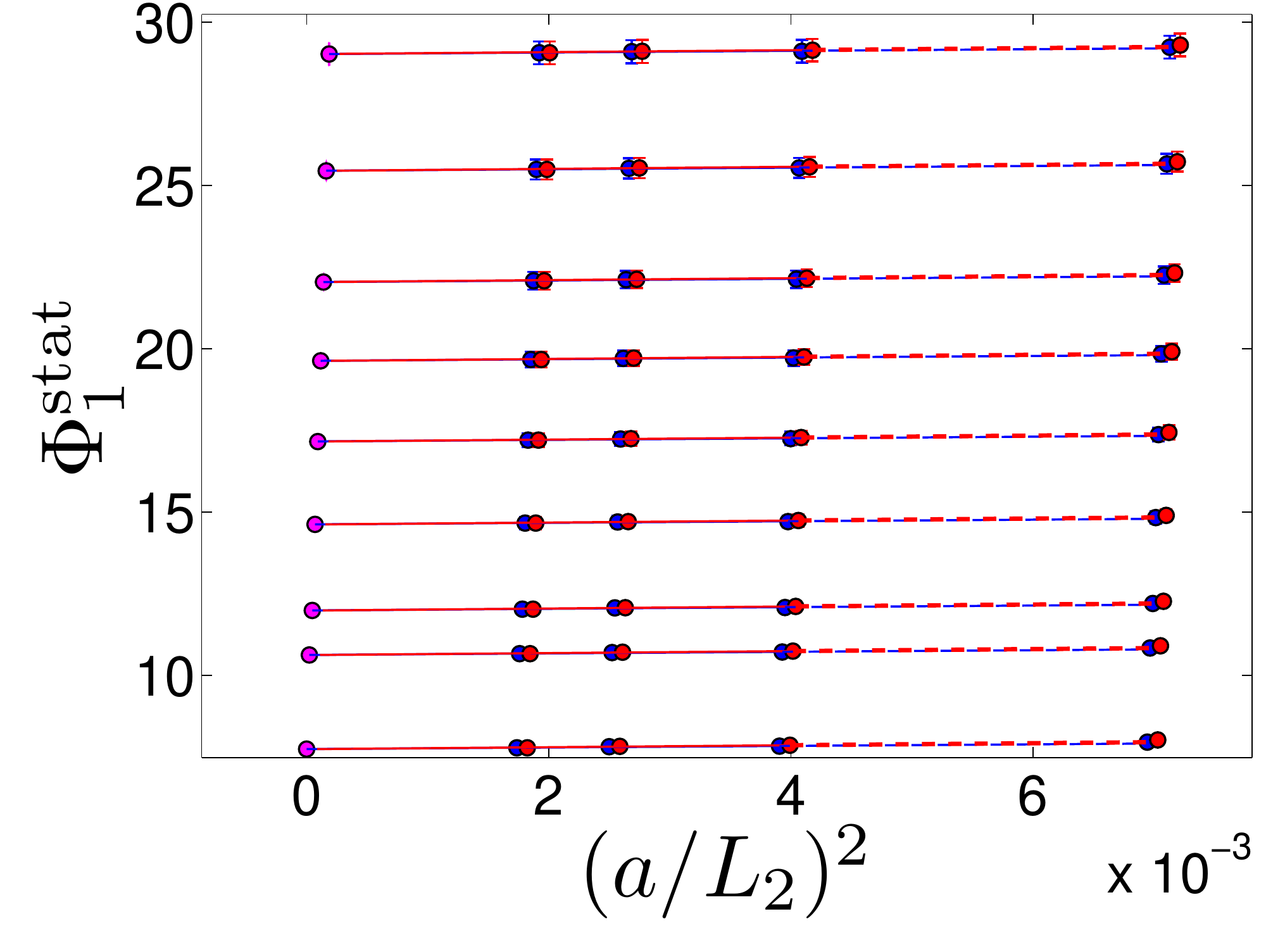}
\includegraphics[width=0.23\textwidth]{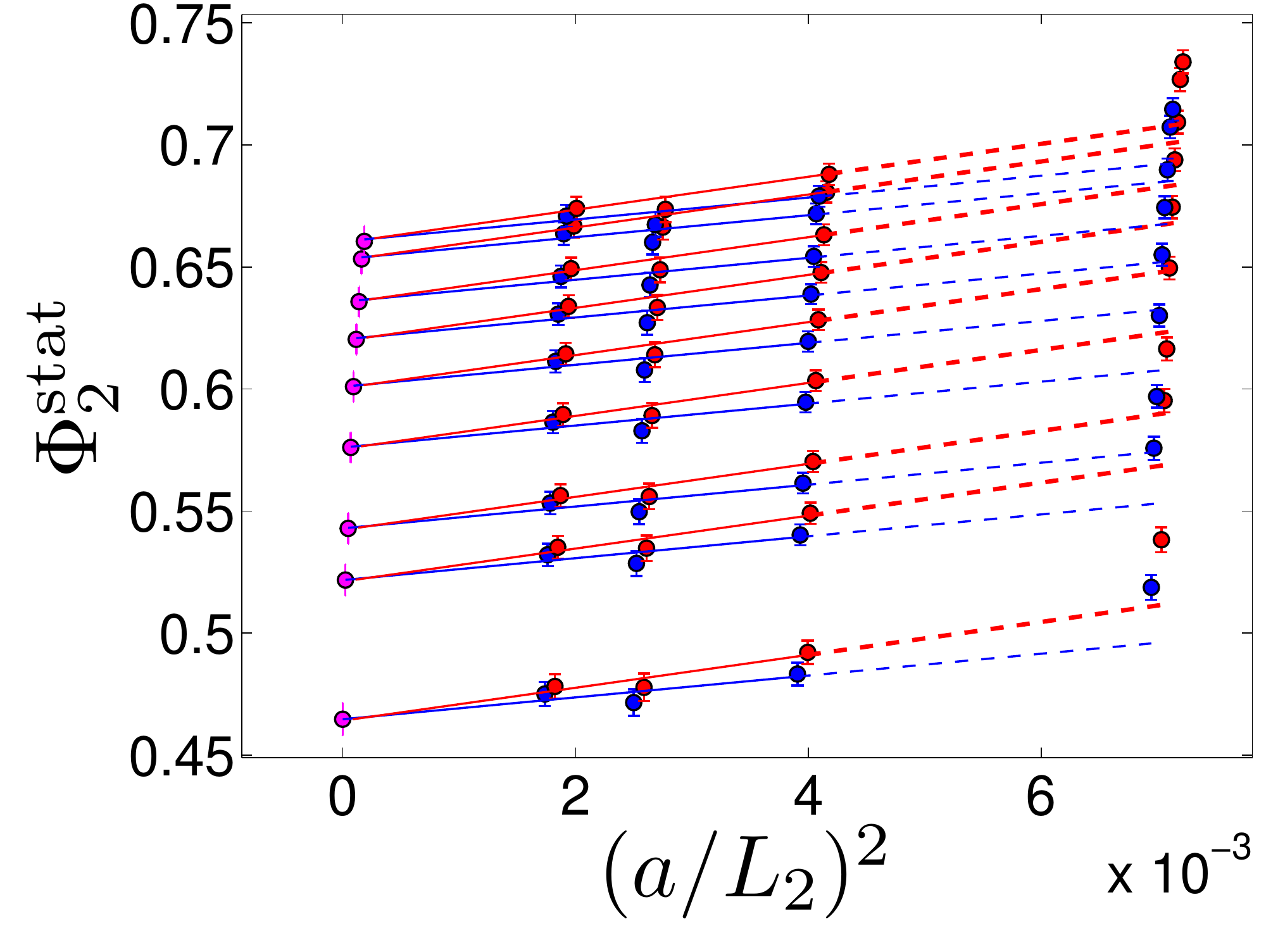}
\vspace{-1.25cm}
\caption{\sl%
Continuum extrapolation of the static approximation of $\Phi_1$ and $\Phi_2$ 
in the volume of extent $L_2$. Red (blue) symbols refer to the HYP1 (HYP2)
discretization of the static propagator.
}\label{fig:Phi12L2}
\vspace{-0.75cm}
\end{center}
\end{figure}

When the effective theory is simulated in the same physical volume
($S_2$ in \Fig{fig:strat}), a set of matching conditions for
$\lim_{a\to 0}\Phi_i^{\rm QCD}(L_1,M,a)$,
\[
\Phi^{\rm QCD}(L_1,M)= 
\eta(L_1,a)+\phi(L_1,a)\tilde\omega(M,a)\,,  
\]
is imposed; the r.h.s. represents the heavy quark mass expansion of the 
$\Phi^{\rm QCD}$ at $\Or(1/\mb)$.
Having computed $\eta$ and $\phi$ from these simulations for different 
values of $a$, the matching equations determine the set of parameters 
$\tilde\omega(M,a)$; e.g., in the simple case of the static meson mass, and 
up to a kinematic constant, $\eta$ is the static energy, $\phi$ a constant 
and $\omega$ the bare static quark mass.
After step scaling to $L_2=2L_1$, the observables in this volume are 
now obtained, thanks to the parameters $\omega(M,a)$ fixed by the previous 
step, as
\[
\Phi(L_2,M,0)= 
\lim_{a\to 0}\left[\,\eta(L_2,a)+\phi(L_2,a)\tilde\omega(M,a)\,\right]\,,
\]
and the continuum limit can be taken, since the power divergences in HQET
cancel out here.
\Fig{fig:Phi12L2} depicts examples of corresponding continuum 
extrapolations in the static approximation, and the results for observables 
sensitive to the $1/\mb$--corrections are of similar 
quality~\cite{lat10:hqetNf2}.
%
%

Finally, the HQET parameters to be employed in the large volume, 
$L_{\infty}$, are estimated from $S_3$:
\[
\omega(M,a)= 
\phi^{-1}(L_2,a)\left[\,\Phi(L_2,M,0)-\eta(L_2,a)\,\right]\,.
\]
\subsection{Preliminary large-volume results}
\label{Sec_appl_largeV}
\begin{figure}[htb]
\begin{center}
\vspace{-1.25cm}
\includegraphics[width=0.25\textwidth]{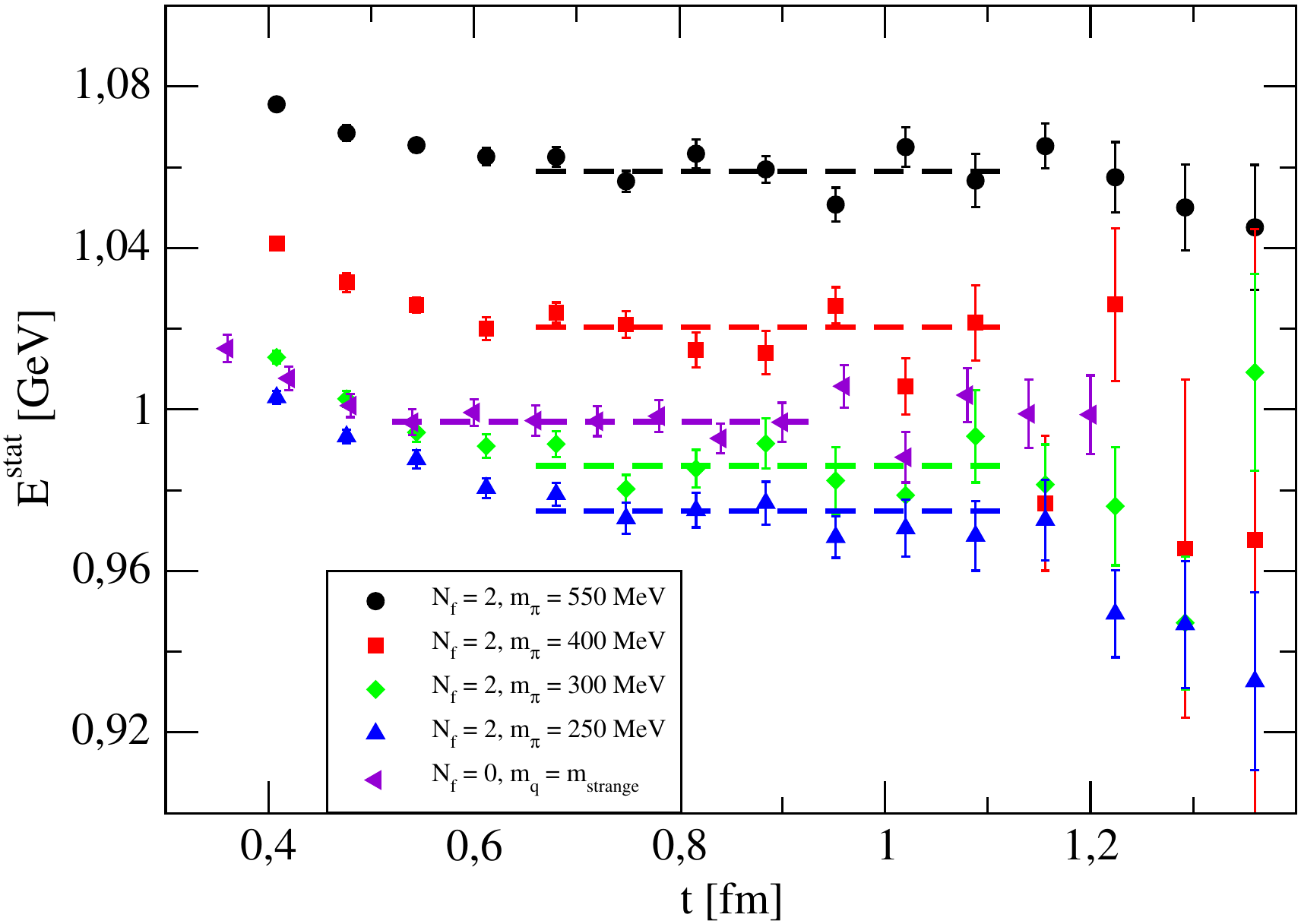}
\includegraphics[width=0.21\textwidth]{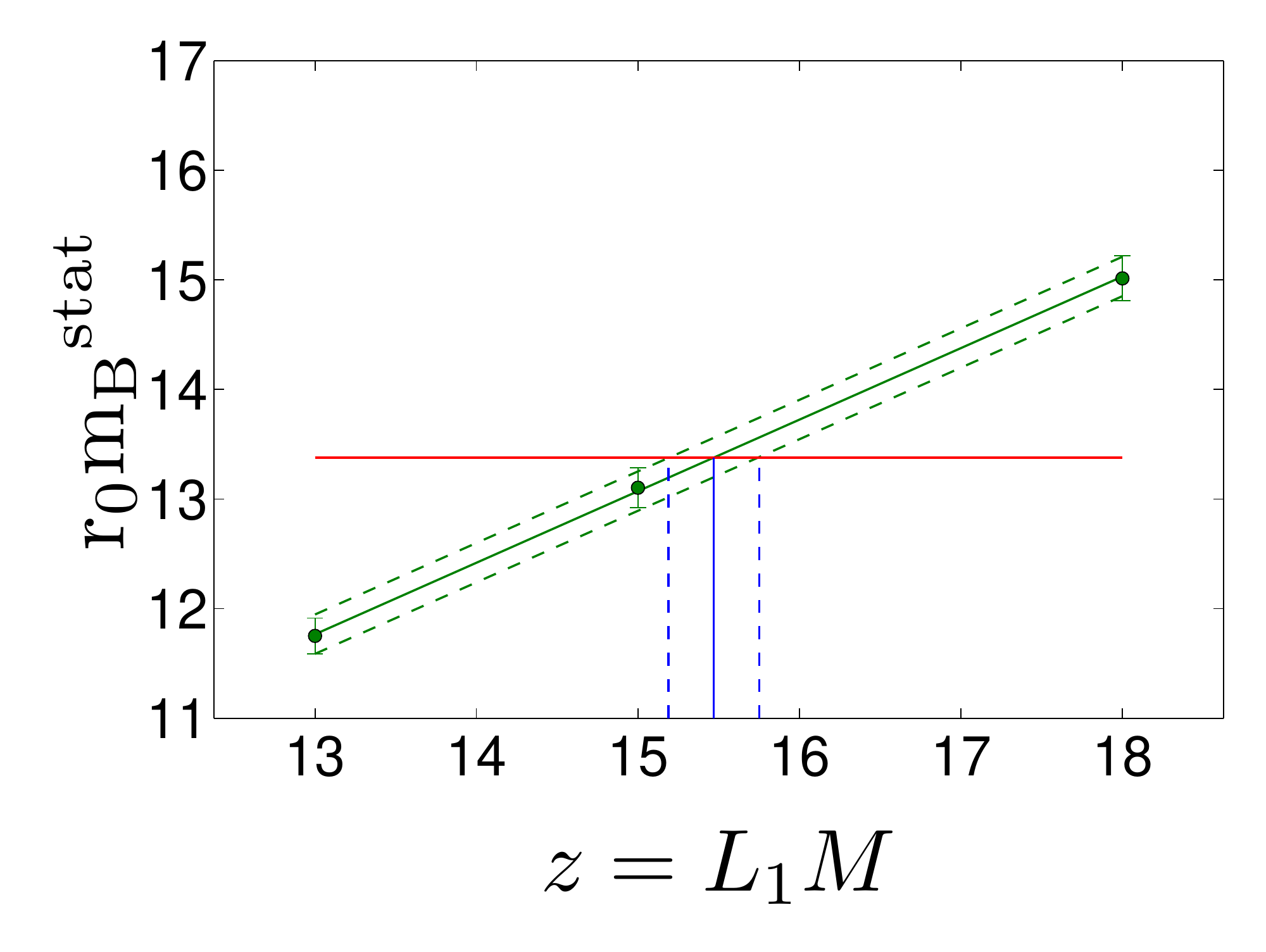}
\vspace{-1.375cm}
\caption{\sl%
Left:
Comparison of plateaux of static energies at $\beta=5.3$ to earlier
quenched results.
Right:
Graphical solution of~(\ref{mastereq_E}) in static approximation; 
$M\equiv\Mh$ is the RGI heavy quark mass.
}\label{fig:Mb_stat}
\vspace{-1.25cm}
\end{center}
\end{figure}

To apply the non-perturbative matching results to calculate the b-quark 
mass, we write down the HQET expansion (to first order in $1/\mb$) of $\mB$ 
in terms of HQET parameters and energies as
\beq
\mB=
m_{\rm bare}+E^{\rm stat}
+\omega_{\rm kin}E^{\rm kin}
+\omega_{\rm spin}E^{\rm spin}\,.
\label{mastereq_E}
\eeq
HQET energies and matrix elements have been extracted from measurements on a 
subset of configuration ensembles produced within CLS~\cite{community:CLS} 
solving the Generalized Eigenvalue Problem \cite{HQET:gevp}, which allows 
for a clean quantification of systematic errors from excited state 
contaminations.
So far, only a single lattice spacing $a\approx 0.07\,\Fm$ ($\beta=5.3$) has 
been analyzed so that the size of discretization effects can not be assessed 
yet.
\Fig{fig:Mb_stat} shows elements of the computations in $L_{\infty}$.
%
%

Upon chiral extrapolation in the light quark mass and including a 
conservative uncertainty in the lattice scale 
($r_{0}=0.475(25)\,\Fm$~\cite{lat10:bjoern}), we quote as our preliminary 
result for the b-quark's mass in HQET at $\Or(1/\mb)$ for the $\nf=2$ 
theory: 
\[
\mbbMS(\mbbar)=
4.276(25)_{r_0}(50)_{{\rm stat}+{\rm renorm}}(?)_a\,\GeV\,.
\]
The first error states the scale uncertainty, while the second covers the 
statistical errors of HQET energies, the chiral extrapolation uncertainty 
and the error on the quark mass renormalization entering the small-volume 
QCD part of the computation ($S_1$).
More details are found in~\cite{lat10:hqetNf2}.

For comparison, we cite the previous $\nf=0$ HQET result
$\mbbMS(\mbbar)=4.320(40)_{r_0}(48)\,\GeV$ by our 
collaboration~\cite{HQET:mb1m} and the recent sum-rule determination,
$\mbbMS(\mbbar)=4.163(16)\,\GeV$ \cite{mbottom:karlsruhe}.
\section{Outlook}
\label{Sec_outl}
The non-perturbative treatment of HQET including $1/\mb$--terms can lead
to results with unprecedented precision for B-physics on the lattice. 
It also greatly improves our confidence in the use of the effective theory.
Our project to extract from $\nf=2$ lattice simulations relevant quantities 
for B-phenomenology within HQET is well advanced.
While the non-perturbative matching of HQET with QCD through small-volume 
simulations is almost done, the evaluation of HQET energies and matrix 
elements has started recently on the CLS ensembles, but still awaits a 
better control of the cutoff effects.
Our first results for the b-quark mass in HQET at $\Or(1/\mb)$ are 
promising, and further applications of the once determined HQET parameters 
to calculate the B-meson decay constant, the spectrum of heavy-light mesons 
and the form factors of the ${\rm B}\to\pi$ semi-leptonic decay are 
expected in the future.
\section*{Acknowledgments}
%
%
I am indebted to my colleagues in CLS and ALPHA for a fruitful 
collaboration, in particular to B.~Blossier, J.~Bulava, M.~{Della Morte}, 
M.~Donnellan, P.~Fritzsch, N.~Garron, G.~von Hippel, B.~Leder, N.~Tantalo, 
H.~Simma and R.~Sommer in the context of our common project on B-physics 
phenomenology from $\nf=2$ lattice simulations.
We acknowledge support by the Deutsche Forschungsgemeinschaft in the 
SFB/TR~09-03, ``Computational Particle Physics'', and under grant
HE~4517/2-1, as well as by the European Community through EU Contract 
No.~MRTN-CT-2006-035482, ``FLAVIAnet''.
We thank CLS for the joint production and use of gauge 
configurations~\cite{community:CLS}.
Our simulations are performed on BlueGene, PC clusters and apeNEXT of the 
John von Neumann Institute for Computing at FZ J\"ulich, at HLRN, Berlin, 
DESY, Zeuthen, and INFN, University of Rome ``Tor Vergata''.
We thankfully acknowledge the computer resources and support provided by 
these institutions.
%
%
\bibliography{lattice_ALPHA}
\bibliographystyle{h-elsevier3.bst}
\end{document}